\documentclass[aps,prl,twocolumn,showpacs,groupedaddress]{revtex4}  

\usepackage{graphicx}  
\usepackage{dcolumn}   
\usepackage{bm}        
\usepackage{amssymb}   
\usepackage{amsmath}

\begin{document}


\title{The magnetoresistance tensor of $\rm{La_{0.8}Sr_{0.2}MnO_3}$}

\author{Y. Bason$^1$}
\email{basony@mail.biu.ac.il}
\author{J. Hoffman$^2$, C. H. Ahn$^2$}
\author{L. Klein$^1$}
\affiliation{$^1$Department of Physics, Nano-magnetism Research Center, Institute of Nanotechnology and Advanced Materials, Bar-Ilan University, Ramat-Gan 52900, Israel}

\affiliation{$^2$Department of Applied Physics, Yale University, New
Haven, Connecticut 06520-8284, USA}

\date{\today}

\begin{abstract}
We measure the temperature dependence of the anisotropic magnetoresistance (AMR) and the planar Hall effect (PHE) in c-axis oriented epitaxial thin films of $\rm{La_{0.8}Sr_{0.2}MnO_3}$, for different current directions relative to the crystal axes, and show that both AMR and PHE depend strongly on current orientation. We determine a magnetoresistance tensor, extracted to $\rm{4^{th}}$ order, which reflects the crystal symmetry and provides a comprehensive description of the data. We extend the applicability of the extracted tensor by determining the bi-axial magnetocrystalline anisotropy in our samples.
\end{abstract}


\maketitle

The interplay between spin polarized current and magnetic moments gives rise to intriguing phenomena which have led to the emergence of the field of spintronics \cite{spintronics}. In most cases, the materials used for studying these phenomena have been amorphous alloys of 3d itinerant ferromagnets (e.g., permalloy), while much less is known about the behavior in materials which are crystalline and more complicated. Manganites, which are magnetic perovskites,  serve as a good example for such a system. As we will show, elucidating these phenomena in this material system provides tools for better theoretical understanding of spintronics phenomena and reveals opportunities for novel device applications.

The magnetotransport properties of manganites known for their colossal magnetoresistance have been studied quite extensively; nevertheless, along numerous studies devoted to elucidating the role of the magnitude of the magnetization, relatively few reports have addressed the role of the \emph{orientation} of the magnetization, which is known to affect both the longitudinal resistivity $\rho_{long}$
(anisotropic magnetoresistance effect - AMR) and transverse
resistivity $\rho_{trans}$ (planar Hall effect - PHE).

For conductors that are amorphous magnetic films, the dependence of $\rho_{long}$ and $\rho_{trans}$ on the magnetic orientation is given by:

\begin{equation}
\rho_{long} = \rho_{\perp} + (\rho_ {\parallel} - \rho_{\perp}) \cos^{2} \varphi
\label{Eq:Rho_xx_old}
\end{equation}
and
\begin{equation}
\rho_{trans} = (\rho_{\parallel} - \rho_{\perp}) \sin \varphi \cos\varphi
\label{Eq:Rho_xy_old}
\end{equation}
where $\varphi$ is the angle between the current \textbf{J} and the magnetization \textbf{M} and $\rho_{\parallel}$ and $\rho_{\perp}$ are the resistivities parallel and perpendicular to \textbf{M}, respectively \cite{phe1,amr1}.
Eqs. \ref{Eq:Rho_xx_old} and \ref{Eq:Rho_xy_old} are not expected to apply to crystalline conductors, as they are independent of the crystal axes \cite{Doring}.
Nevertheless, they have been used to describe AMR and PHE in epitaxial films \cite{PHE_Magnetite,PHE_GaMnAs,PHE_LSMO,Deviations_Manganites}; qualitative and quantitative deviations were occasionally attributed to extrinsic effects.

Here, we quantitatively identify the crystalline contributions to AMR and PHE in epitaxial films of La$_{0.8}$Sr$_{0.2}$MnO$_{3}$ (LSMO) and replace Eqs. \ref{Eq:Rho_xx_old} and \ref{Eq:Rho_xy_old} with equations that provide a comprehensive description of the magnetotransport properties of LSMO. The equations are derived by expanding the resistivity tensor to $4^{th}$ order and keeping terms consistent with the crystal symmetry.

AMR and PHE in manganites constitute an important aspect of their magnetotransport properties; hence, quantitative determination of these effects is essential for comprehensive understanding of the interplay between magnetism and transport in this class of materials. In addition, when the dependence of AMR and PHE on local magnetic configurations is known, the two effects can be used as a powerful tool for probing and tracking static and dynamic magnetic configurations in patterned structures. Moreover, as the magnitude of the AMR and PHE changes dramatically with current direction, the elucidation of the appropriate equations is crucial for designing novel devices with optimal properties that are based on these phenomena.

Our samples are epitaxial thin films ($\rm{\sim40\ nm}$) of LSMO with a Curie temperature ($\rm{T_c}$) of $\sim 290 \ {\rm K}$ grown on cubic single crystal [001] $\rm{SrTiO_3}$ substrates using off-axis magnetron
sputtering. $\theta-2\theta$ x-ray diffraction reveals c-axis
oriented growth (in the pseudocubic frame), with an out-of-plane lattice constant
of $\rm{\sim 0.3876\ nm}$, and an in-plane lattice constant of $\rm{\sim 0.3903\ nm}$, consistent with coherently strained films. No impurity
phases are detected. Rocking curves taken around the 001 and 002 reflections
have a typical full width at half maximum of $0.05^o$. The film surfaces have been characterized using atomic force microscopy, which
shows a typical root-mean-square surface roughness of $\rm{\sim 0.2\ nm}$. The samples were patterned using photolithography to create 7 patterns on the same substrate. Each pattern has its current path at a different angle $\theta$ relative to the [100] direction ($\theta=0^\circ, 15^\circ, 30^\circ, 45^\circ,
60^\circ, 75^\circ, 90^\circ$), with electrical leads that allow for AMR and PHE measurements.

\begin{figure*}
\begin{center}
\includegraphics[scale=0.4]{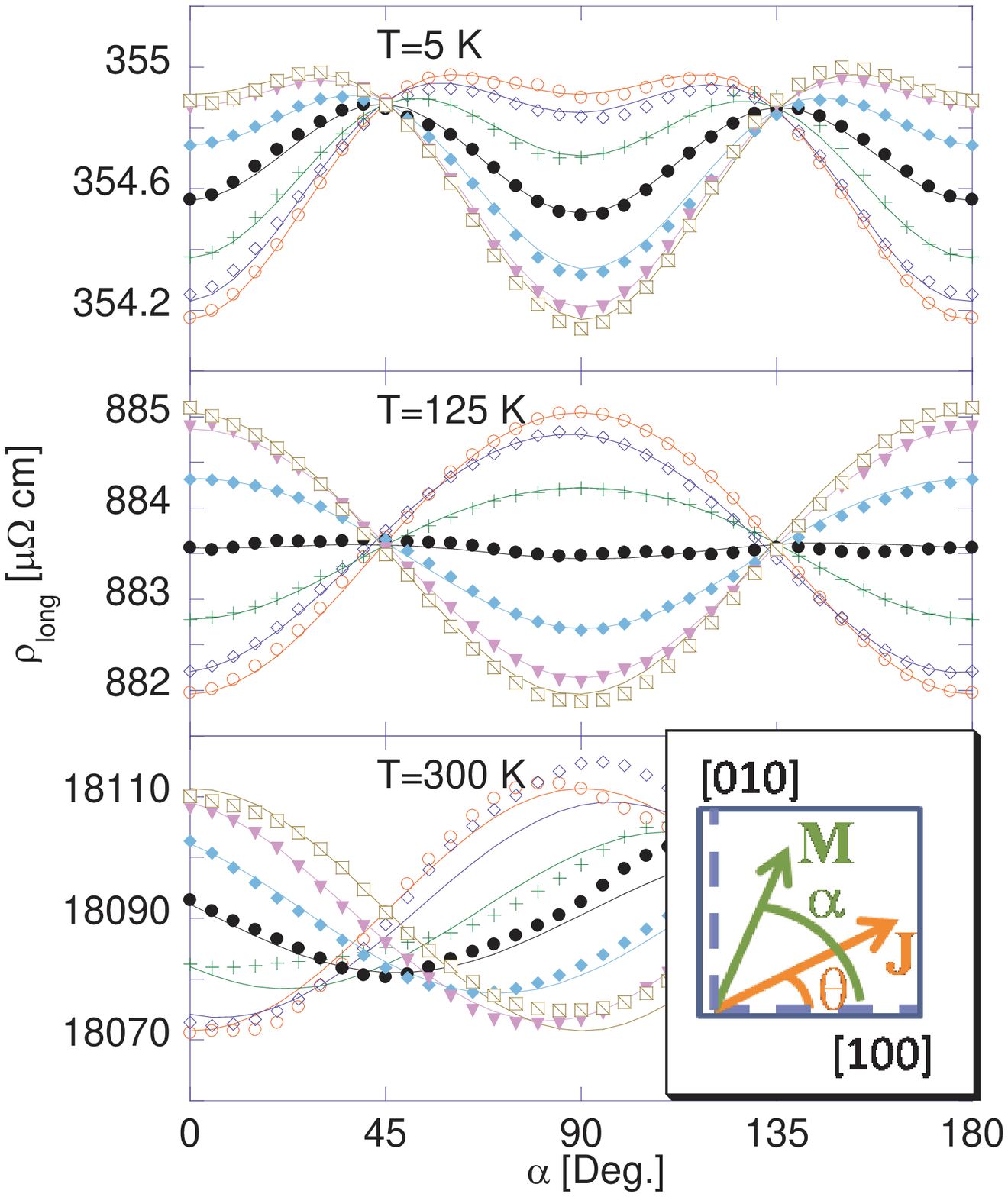}
\includegraphics[scale=0.4]{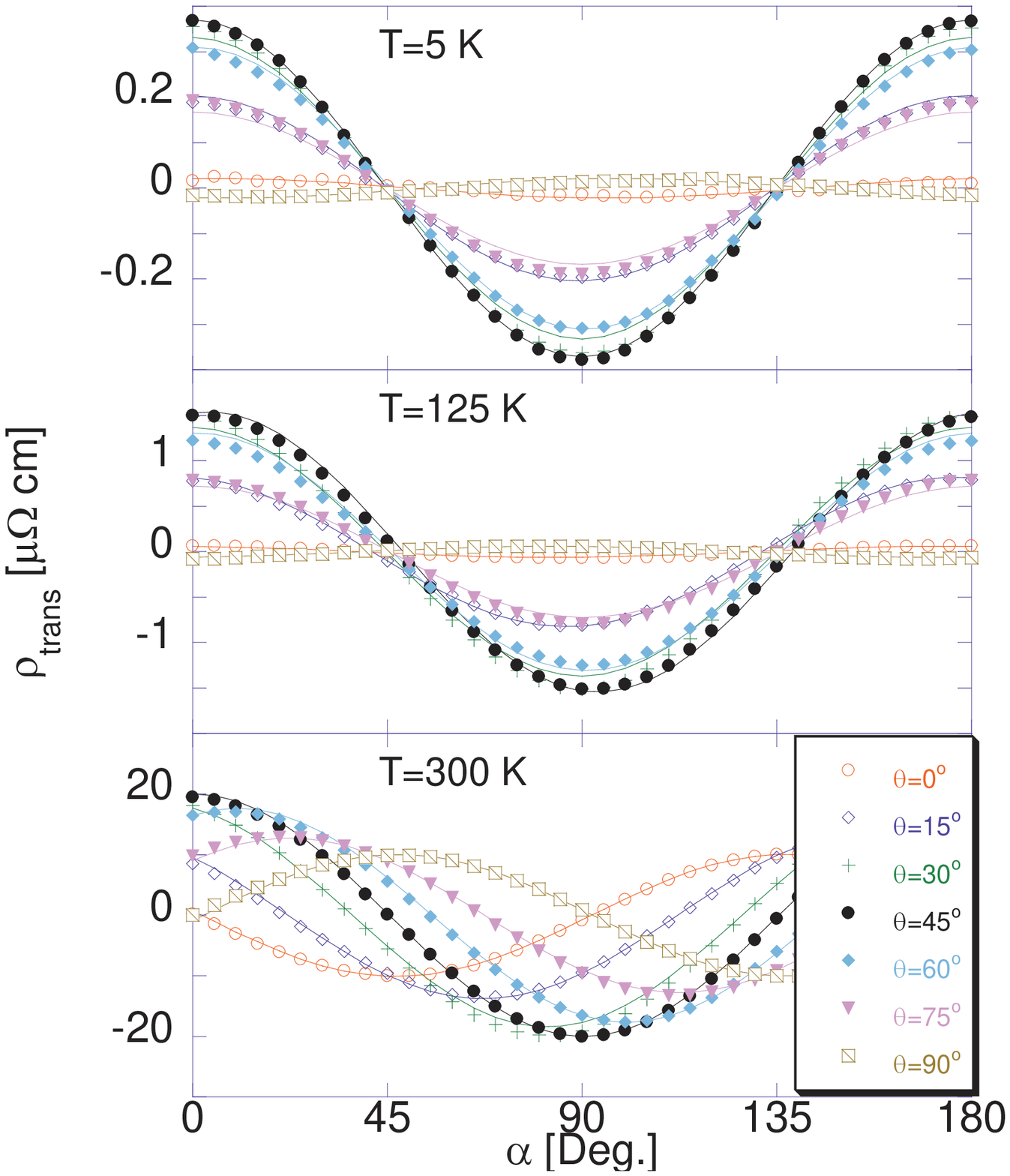}
\end{center}
\caption{\label{Fig:layout} Longitudinal resistivity $\rho_{long}$ (left) and transverse resistivity $\rho_{trans}$ (right) vs. $\alpha$, the angle between the magnetization and [100], for different angles $\theta$ (the angle between the current direction and [100]) at different temperatures with an applied magnetic field of 4 T. The solid lines are fits to Eqs. \ref{Eq:Rho_xx} and \ref{Eq:Rho_xy}. Inset: Sketch of the relative orientations of the current density \textbf{J}, magnetization \textbf{M}, and the crystallographic axes.
}
\end{figure*}

Fig. \ref{Fig:layout} presents $\rho_{long}$ and $\rho_{trans}$ data obtained by applying a field of H=4 T in the film plane and rotating the sample around the [001] axis. The figure shows the data for all seven patterns at T=5, 125 and 300 K. At $\rm{T=300\ K}$ both $\rho_{long}$ and $\rho_{trans}$ seem to behave according to Eqs. \ref{Eq:Rho_xx_old} and \ref{Eq:Rho_xy_old}. However, contrary to these equations, the amplitude of $\rho_{long}$ differs from the amplitude of $\rho_{trans}$; moreover, they both change with $\theta$, the angle between \textbf{J} and [100].

The discrepancies increase as the temperature decreases, and at T=125 K the variations in the amplitudes for measurements taken for different $\theta$ increase. Furthermore, the location of the extremal points are dominated by $\alpha$, the angle between \textbf{M} and [100]. At $\rm{T=5\ K}$, the deviations are even more evident as the AMR measurements are no longer described with a sinusoidal curve. All these observations clearly indicate the need for a higher order tensor to adequately describe the magnetotransport behavior of LSMO.

The resistivity tensor in a magnetic conductor depends on the direction
cosines, $\alpha_i$, of the magnetization vector, and can be
expressed as a series expansion of powers of the $\alpha_i$ \cite{PHE_Tensor},
giving:

\begin{eqnarray}
\nonumber
\rho_{ij}(\alpha)=\sum_{k,l,m...=1}^3(a_{ij} + a_{kij}\alpha_k +
a_{klij}\alpha_k\alpha_l +\\
+a_{klmij}\alpha_k\alpha_l\alpha_m +
a_{klmnij}\alpha_k\alpha_l\alpha_m\alpha_n+...)
\label{eq_rho}
\end{eqnarray}
where $i,j=1,2,3$ and the $a$'s are expansion coefficients. As usual $\rho_{ij}(\alpha)=\rho_{ij}^s(\alpha)+\rho_{ij}^a(\alpha)$ where $\rho_{ij}^s$ and $\rho_{ij}^a$ are symmetric and antisymmetric tensors, respectively. As both AMR and PHE are symmetric, we use only $\rho_{ij}^s$ for their expression. As we are interested only in the in-plane properties, we use the tensor expansion for crystals with m3m cubic-crystal structure \cite{SymmetryMagnetism}.
The $4^{th}$ order symmetric resistivity tensor $\rho^s$ for this class of materials in the xy plane (as \textbf{M}, \textbf{J} and the measurements are all in the plane of the film) is given by:

\begin{equation}
\rho^s=\left(
         \begin{array}{ccc}
           C'+C_1'\alpha_1^2+C_2'\alpha_1^4 & C_4'\alpha_1\alpha_2 &  \\
           C_4'\alpha_1\alpha_2 & C'+C_1'\alpha_2^2+C_2'\alpha_2^4
         \end{array}
       \right).
\label{rho_tensor}
\end{equation}
When \textbf{J} is along $\theta$ we obtain:
\begin{equation}
\rho_{long}=A\cos(2\alpha-2\theta)+B\cos(2\alpha+2\theta)+C\cos(4\alpha)+D
\label{Eq:Rho_xx}
\end{equation}
and
\begin{equation}
\rho_{trans}=A\sin(2\alpha-2\theta)-B\sin(2\alpha+2\theta)
\label{Eq:Rho_xy}
\end{equation}
with:

$A=\left(C_1'+C_2'+C_4'\right)/4$

$B=\left(C_1'+C_2'-C_4'\right)/4$

$C=C_2'/8$

$D=C'+C_1'/2+3C_2'/8$

Equations \ref{Eq:Rho_xx} and \ref{Eq:Rho_xy}, which take into account the crystal symmetry, have 4 independent parameters ($A$, $B$, $C$ and $D$) with which we fit (as shown in Figure \ref{Fig:layout}) at any given temperature and magnetic field a set of 14 different curves (7 AMR curves and 7 PHE curves).

The parameter $A$ is a coefficient of a term describing a non-crystalline contribution since ($\alpha - \theta$) is the angle between \textbf{M} and \textbf{J} irrespective of their orientation relative to the crystal axes. On the other hand, the parameters $B$ and $C$ are coefficients of terms that depend on the orientation of \textbf{M} and/or \textbf{J} relative to the crystal axes.

We note that adding the terms with the coefficient $B$ (in both Eq. \ref{Eq:Rho_xx} and \ref{Eq:Rho_xy}) to the "$A$" term changes only the amplitude and the phase of the signal compared to Eqs. \ref{Eq:Rho_xx_old} and \ref{Eq:Rho_xy_old}: Eq. \ref{Eq:Rho_xx} can be written (for C=0) as:
\begin{equation}
\rho_{long}=E\cos(2\alpha-\phi_{long})+D
\label{Eq:Rho_xx_2}
\end{equation}
where $E^2=A^2+B^2+2AB\cos4\theta$ and $\sin\phi_{long}=\frac{A-B}{E}\sin(2\theta)$; and Eq. \ref{Eq:Rho_xy} can be written as:
\begin{equation}
\rho_{trans}=F\sin(2\alpha-\phi_{trans})
\label{Eq:Rho_xy_2}
\end{equation}
where $F^2=A^2+B^2-2AB\cos4\theta$ and $\sin\phi_{trans}=\frac{A+B}{F}\sin2\theta$.
The amplitude of $\rho_{trans}(\alpha)$, $F$, varies with $\theta$ between a maximal value of $|A+B|$ for $\theta=\pm 45^\circ$ and a minimal value of $|A-B|$ for $\theta=0,\pm 90^\circ$. On the other hand, the amplitude of $\rho_{long}(\alpha)$, $E$, obtains its maximal value $|A+B|$ at $\theta=0,\pm 90^\circ$ and its minimal value $|A-B|$ at $\theta=\pm45^\circ$.
When the $C$ term is added it does not affect $\rho_{trans}$; however, $\rho_{long}$ behaves qualitatively differently.

We thus observe that the current direction affects quite
dramatically the amplitude of the effect. At 125 K, for instance, the PHE amplitude for current at $45^\circ$ relative to [100] is more than 20 times larger than the PHE for current parallel to
[100]. This means that appropriate selection of the current direction
that takes into consideration crystalline effects is important for designing devices that use the PHE for magnetic sensor or magnetic memory applications \cite{PHE_MRAM}.

\begin{figure*}
\begin{center}
\includegraphics[scale=0.5]{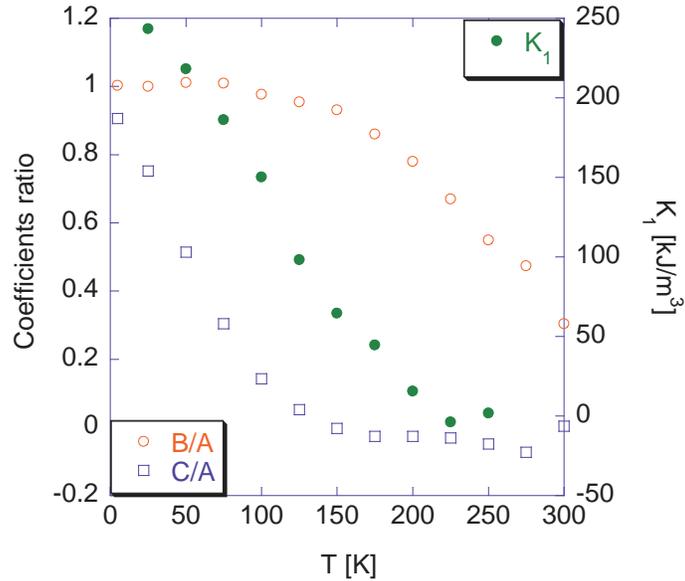}
\end{center}
\caption{\label{Fig:Factors} The ratios of the coefficients from Eqs. \ref{Eq:Rho_xx} and \ref{Eq:Rho_xy} (B/A and C/A) (left axis) and the coefficient $\rm{K_1}$ from Eq. \ref{Eq:energy} (right axis) as a function of temperature.}
\end{figure*}

Figure \ref{Fig:Factors} presents the
temperature dependence of $B/A$ and $C/A$. Close to $\rm{T_c}$ both $B$ and $C$ are negligible relative to $A$; therefore, AMR and PHE measurements appear to fit Eqs. \ref{Eq:Rho_xx_old} and \ref{Eq:Rho_xy_old}. At intermediate temperatures where $C$ is still much smaller than $A$ (while $B$ and $A$ are of the same order), the signal remains sinusoidal, although its deviation from Eqs. \ref{Eq:Rho_xx_old} and \ref{Eq:Rho_xy_old} becomes quite evident. At low temperatures, $C$ is on the order of $B$, and the AMR signal is no longer sinusoidal.

\begin{figure*}
\begin{center}
\includegraphics[scale=0.4]{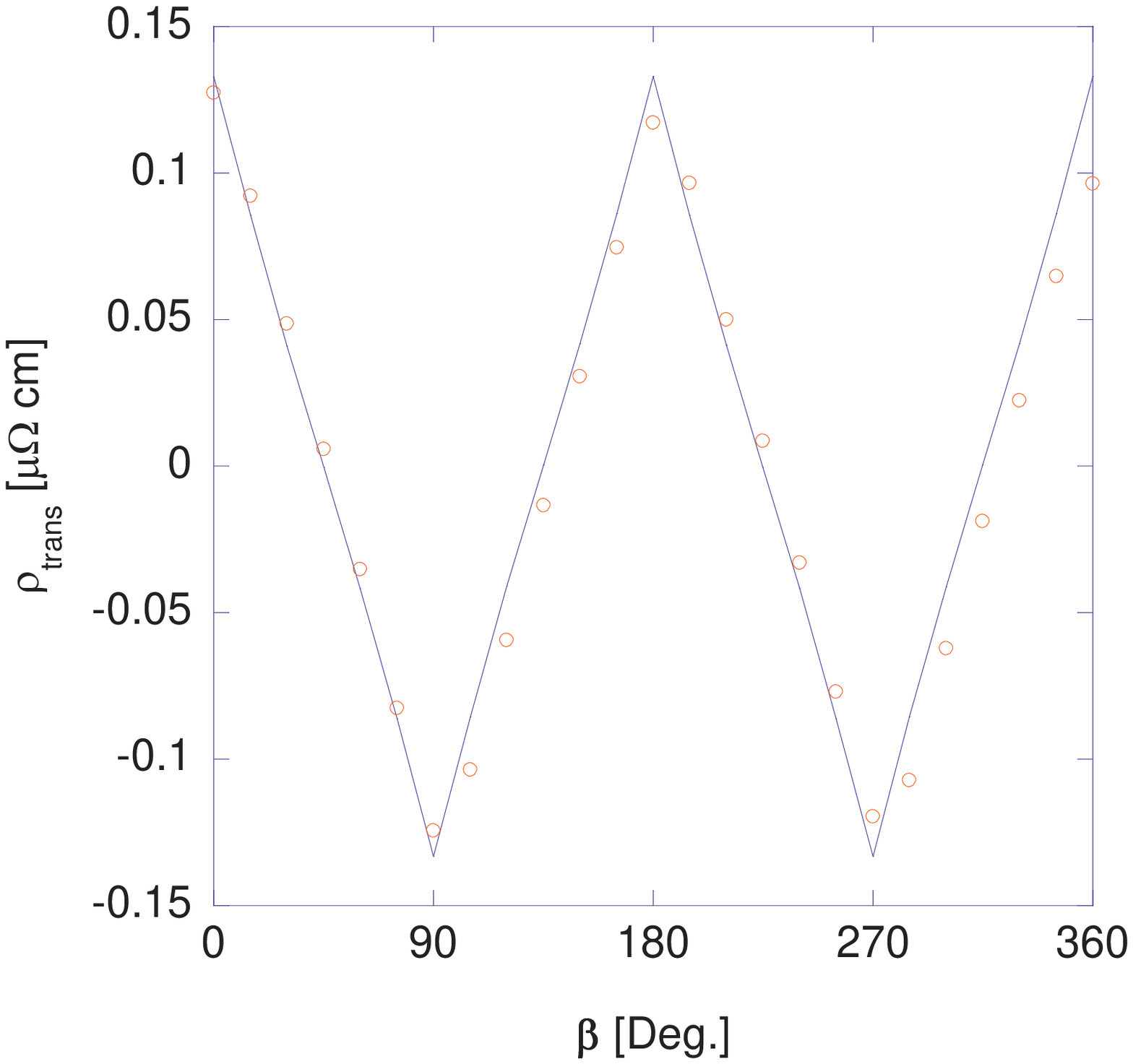}
\includegraphics[scale=0.4]{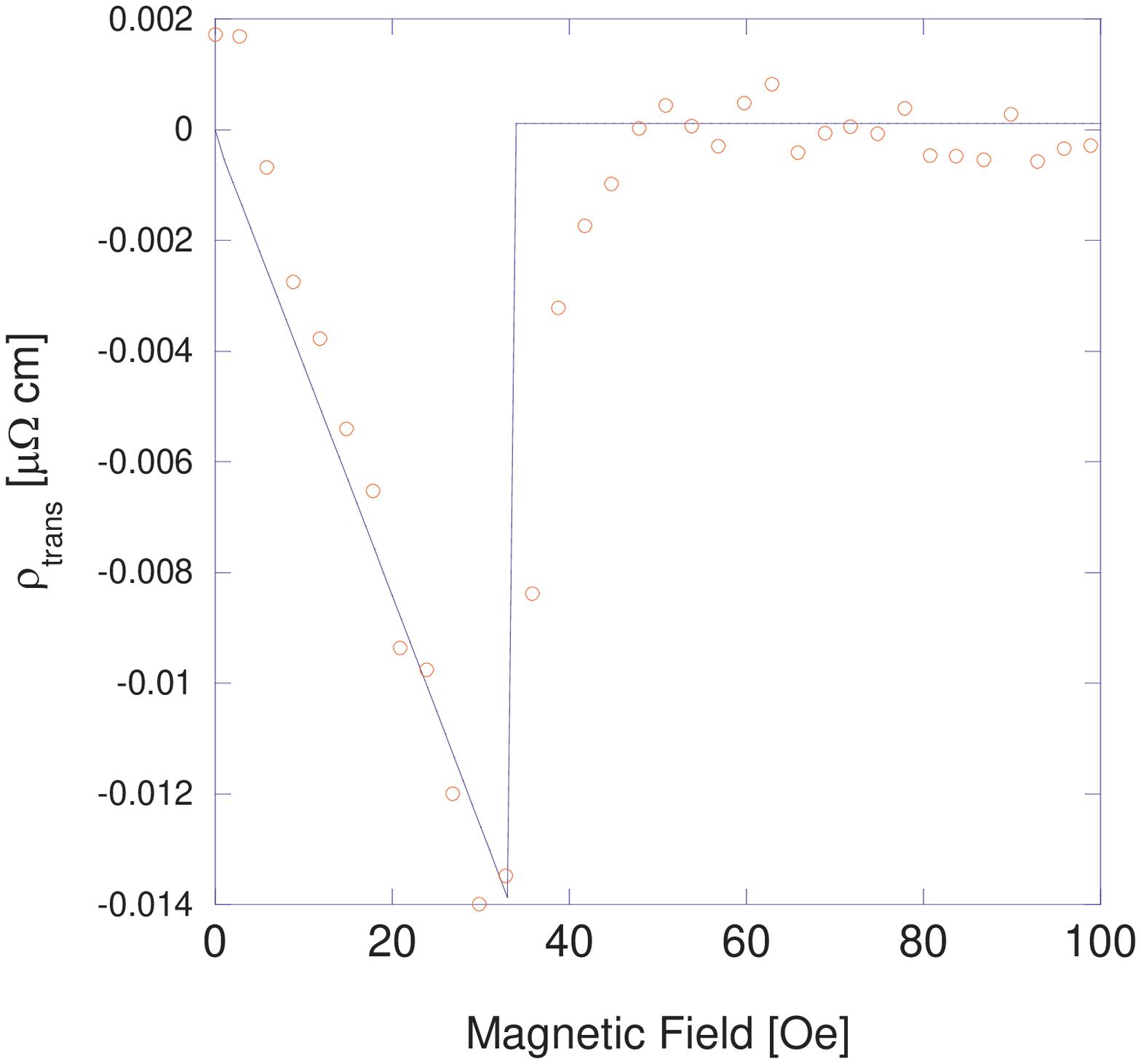}
\end{center}
\caption{\label{Fig:PHELowField} Left: PHE signal as a function of $\beta$, the magnetic field direction relative to the [100] (H=500 Oe and T=50 K). The line is a fit to Eq. \ref{Eq:Rho_xy} with $\alpha$ extracted using Eq. \ref{Eq:energy}. Right: PHE as a function of magnetic field. The sample is prepared with \textbf{M} along [100], and the field is applied along [$1 \bar{1} 0$]. The line is a fit using Eq. \ref{Eq:energy}.}
\end{figure*}

When AMR and PHE measurements are performed with low applied fields, \textbf{M} is no longer parallel to \textbf{H}, due to intrinsic magnetocrystalline anisotropy. Our LSMO films exhibit bi-axial magnetocrystalline anisotropy with easy axes along $\langle110\rangle$ directions, a manifestation of in-plane cubic symmetry. When a field \textbf{H} is applied, the total free energy consists of the magnetocrystalline anisotropy energy and the Zeeman energy:

\begin{equation}
E=\frac{K_1}{4}\cos^2 2\alpha-MH\cos(\alpha-\beta)
\label{Eq:energy}
\end{equation}
where $K_1$ is the magnetocrystalline anisotropy energy and $\beta$ is the angle between \textbf{H} and [100]. The first term is responsible for the bi-axial magnetocrystalline anisotropy with easy axes along $\alpha=\pm\frac{\pi}{4}$ and
$\alpha=\pm\frac{3\pi}{4}$. We have determined the value of $K_1$ at various temperatures (see Fig. \ref{Fig:Factors}) by switching the magnetization between the two easy axes (see Fig. \ref{Fig:PHELowField}). The extracted value of $K_1$ allows us by using Eqs. \ref{Eq:Rho_xx}, \ref{Eq:Rho_xy} and \ref{Eq:energy} to fit the AMR and PHE data obtained with relatively low applied fields (e.g., 500 Oe), where \textbf{M} does not follow \textbf{H} (see Fig. \ref{Fig:PHELowField}).

In summary, we have expanded the magnetoresistance tensor to $\rm{4^{th}}$ order keeping terms consistent with the symmetry of epitaxial films of LSMO and derived equations that provide a comprehensive description of AMR and PHE in LSMO films in a wide range of temperatures. The results shed new light on the interplay between magnetism and electrical transport in this class of materials and may serve as a basis for further study of the microscopic origin of magnetotransport properties of LSMO and other manganites. The results contribute to the ability to monitor magnetic configurations via magnetotransport properties, a feature of particular importance in studying nano-structures, and will facilitate the design of novel devices that use AMR and PHE.

We acknowledge useful discussions with E. Kogan. L.K. acknowledges support by the Israel Science Foundation founded by the Israel Academy of Sciences and Humanities. Work at Yale supported by NSF MRSEC DMR 0520495, DMR 0705799, NRI, ONR, and the Packard Foundation.

\begin{thebibliography}{99}
\bibitem{spintronics} G. A. Prinz, Science \textbf{282}, 1660 (1998).
\bibitem{CMR} M. McCormack, S. Jin, T H. Tiefel, R. M. Fleming, J. M. Phillips and R. Ramesh, Appl. Phys. Lett. \textbf{64}, 3045 (1994);
\bibitem{phe1} C. Goldberg and R. E. Davis, Phys. Rev. {\bf 94}, 1121 (1954); F. G. West, J. Appl. Phys. {\bf 34}, 1171 (1963); W. M. Bullis, Phys. Rev. {\bf 109}, 292 (1958).
\bibitem{amr1} T. R. McGuire and R. I. Potter, IEEE Trans. Magn. {\bf 11}, 1018 (1975).
\bibitem{Doring} D\"{o}ring, W., Ann. Physik \textbf(32) (1938) 259.
\bibitem{PHE_Magnetite} X. Jin, R. Ramos, Y. Zhou, C. McEvoy, and I. V. Shvets, J. Appl. Phys. {\bf 99}, 08C509 (2006);
\bibitem{PHE_GaMnAs} H. X. Tang, R. K. Kawakami, D. D. Awschalom, and M. L. Roukes, PRL \textbf{90},107201 (2003);
\bibitem{PHE_LSMO} Y. Bason, L. Klein, J.-B. Yau, X. Hong, and C. H. Ahn, Appl. Phys. Lett. \textbf{84}, 2593 (2004).
\bibitem{Deviations_Manganites} I.C. Infante, V. Laukhin, F. S$\rm{\acute{a}}$nchez, J. Fontcuberta, Materials Science and Engineering B \textbf{126} (2006) 283-286;
\bibitem{PHE_Tensor} T. T. Chen, V. A. Marsocii, Physica \textbf{59} (1972) 498-509.
\bibitem{SymmetryMagnetism} Birss, R. R., Symmetry and Magnetizm, North-Holland Publ. Comp. (Amsterdam, 1964).
\bibitem{PHE_MRAM} Y. Bason, L. Klein, J.-B. Yau, X. Hong, J. Hoffman, and C. H. Ahn, J. Appl. Phys. \textbf{99}, 08R701 (2006).
\end{thebibliography}
\end{document}